# From Articles to Code: On-Demand Generation of Core Algorithms from Scientific Publications

Cameron S. Movassaghi[†], Amanda Momenzadeh[†], Jesse G. Meyer*

Department of Computational Biomedicine, Cedars Sinai Medical Center, Los Angeles, CA 90048
[†]These authors contributed equally to this work.
*Correspondence to jessegmeyer@gmail.com

**ABSTRACT**
Maintaining software packages imposes significant costs due to dependency management, bug fixes, and versioning. We show that rich method descriptions in scientific publications can serve as standalone specifications for modern large language models (LLMs), enabling on-demand code generation that could supplant human-maintained libraries. We benchmark state-of-the-art models (GPT-o4-mini-high, Gemini Pro 2.5, Claude Sonnet 4) by tasking them with implementing a diverse set of core algorithms drawn from original publications. Our results demonstrate that current LLMs can reliably reproduce package functionality with performance indistinguishable from conventional libraries. These findings foreshadow a paradigm shift toward flexible, on-demand code generation and away from static, human-maintained packages, which will result in reduced maintenance overhead by leveraging published articles as sufficient context for the automated implementation of analytical workflows.

**INTRODUCTION**

Scientific articles serve as the foundation of reproducible computational research, providing detailed, peer-reviewed descriptions of novel algorithms and methodologies. However, the journey from paper to robust, usable software is fraught with challenges. Translating narrative algorithmic insights into production-grade implementations remains a labor-intensive process, frequently hindered by ambiguities or omitted practical details in published work. This gap is widely recognized as a core barrier to scientific transparency and the credibility of computational findings, with reproducibility crises and failed software reimplementations demonstrating the difficulties researchers face in reusing published work.[1,2]

To bridge this divide, software libraries provide powerful abstractions, encapsulating complex scientific and statistical methods behind user-friendly APIs. It is well accepted that code libraries can enable reproducible research[2,3]. While these libraries accelerate research, their maintenance is a monumental task. Key challenges include:
- Prominent and deeply interwoven dependency chains, where updates in foundational packages propagate breaking changes throughout the ecosystem, creating a delicate balance between innovation and stability[4];
- Subtle edge-case bugs that require targeted, labor-intensive fixes, often only identified through extensive community usage and evolving scientific requirements;
- Version mismatches and a lack of standardized compute environments undermine the reproducibility of scientific research.

Recent advances in large language models (LLMs) for code generation, such as OpenAI Codex[5] and DeepMind AlphaCode[6], mark a paradigm shift in how computational methods can be instantiated. These models, trained on billions of lines of code and natural language, can translate natural-language problem descriptions into executable software. In several benchmarks, these models not only accelerate development but also demonstrate competitive performance in real-world tasks, with capabilities advancing rapidly across programming languages and domains. However, success rates vary: while straightforward problems are effectively solved, multi-step scientific workflows and nuanced research tasks remain a challenge for current LLMs[5,6].

When paired with retrieval-augmented generation (RAG) frameworks[7], LLMs can fetch precise algorithmic details or API documentation from curated corpora, including scientific articles and official documentation, at inference time. This approach enables a model to synthesize code directly guided by original research, rather than relying solely on historical training data. By dynamically integrating this external knowledge just-in-time, the distinction between published algorithm and runnable implementation begins to fade, hinting at a new paradigm where articles themselves operate as executable specifications.

In this study, we systematically probe the limits of LLM-driven code synthesis using exclusively literature-sourced descriptions. We present a comprehensive benchmark, tasking several state-of-the-art models with reimplementing core methods from popular Python libraries, guided only by a prompt and the original publication's text (see **Figure 1**). Outputs are rigorously compared against package-standard behaviors across diverse test suites. These results highlight where on-demand, literature-driven generation can succeed, where current models fall short, and assess how close we are to a future in which flexible, article-driven implementations supersede traditional package maintenance.

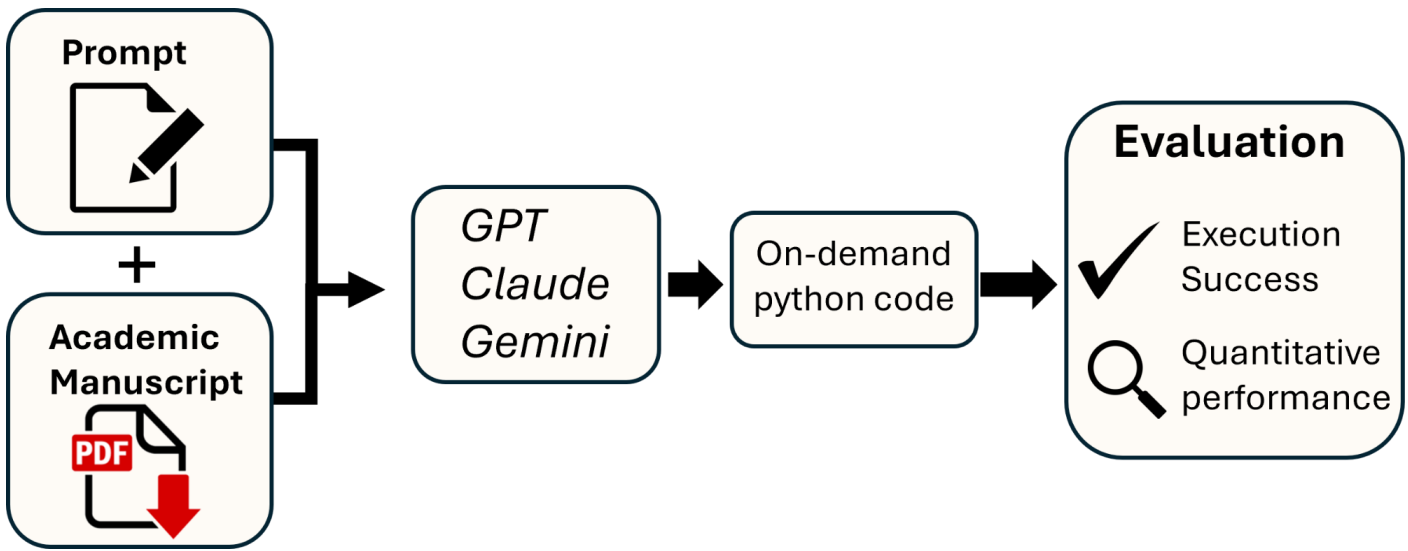

**Figure 1: Concept overview.** Prompts were passed to various models requesting that methods be created based only on the academic manuscript describing the algorithm. The on-demand generated python code was then evaluated with regard to whether it could successfully run at all, and then qualitative and quantitative metrics of performance relative to the intended goal were computed to compare various versions of each method.

**METHODS**

**Code and data availability.** The code, data, and PDF files used in this paper are available from https://github.com/xomicsdatascience/articles-to-code.

**Random Forest.** The original PDF of the paper[8] was attached, and the prompt was: "Forget all our previous chats and forget everything you know about the random forest method for machine learning. Using only the information in the attached PDF, try your hardest to create an implementation of a random forest that exactly matches the algorithm for classification. Your function should work with the iris dataset from sklearn to classify the flower type. You only get one try to do this correctly, so please try very carefully to get the code correct." The code is in RF-compare-models.ipynb.

**Combat.** The original PDF paper[9] was attached, and the first prompt was: "Forget all our previous chats and forget everything you know about the batch correction method called combat. Using only the information in the attached PDF, try your best to create an implementation of combat that exactly matches the math described to correct omics data across batches. It should take an input of a dataframe where columns represent samples and rows represent genes, along with a list of batches for each sample (in the same order as the columns of the dataframe), and then return the corrected dataframe. You only get one try to do this correctly, so please try very carefully to get the code correct." The generated code, including the comparative analysis, is available in combat-compare-models.ipynb.

**Augusta.** The PDF of the paper[10] was attached along with the CSV with data from the relevant Dialogue on Reverse Engineering Assessment and Methods (DREAM) challenge[11] (available from their GitHub repository). The following prompt was used: "Forget anything you know about the method called Augusta, and ignore any previous chats about that method. Use the PDF paper I uploaded and implement their method for gene regulatory network discovery from scratch in Python. Implement the version without requiring motifs from a genome. Use the timeseries of gene expression in the Ecoli_DREAM4.csv file to apply that algorithm to detect a GRN. Your code should produce a GRN as a matrix of gene x gene, where each cell gives a 1 if positive relation or -1 if a negative relation. Run the code yourself and solve any errors that come up until it works." The generated code, including the comparison to their pip package, is available on GitHub in augusta.ipynb.

**Systematic Error Removal by Random Forest (SERRF).** Within the Google Collab environment, a PDF of the paper[12] was attached along with an Excel file with example data. The following prompt was used: "Forget all our previous chats and forget everything you know about the batch correction method called systematic error removal using random forest (SERRF). Using only the information in the attached PDF, try your best to create an implementation of SERRF that exactly matches the math described to correct omics data across batches. It should take a pandas DataFrame as input and return a pandas DataFrame containing the batch-corrected omics data. Test the implementation using the provided example dataset. You only get one try to do this correctly so please try very carefully to get the code correct." The resulting code is available on GitHub in three separate ipynb files: "SERFF from repo url.ipynb", "SERRF from gitingest.ipynb", and "SERRF from pdf.ipynb".

A similar prompt was then attempted with the PDF and the example data using gpt-o4-mini-high: "Forget all our previous chats and forget everything you know about the batch correction method called systematic error removal using random forest (SERRF). Using only the information in the attached PDF, try your best to create an implementation of SERRF that exactly matches the math described to correct omics data across batches. It should take a pandas DataFrame as input and return a pandas DataFrame containing the batch-corrected omics data. Test the implementation using the provided example dataset. You only get one try to do this correctly so please try very carefully to get the code correct. Iteratively try to solve any errors that pop up and run the code yourself to test it until it works. "

After that failed, it was clear that figuring out the multi-index structure of the example dataset without instruction was the problem. A second prompt was tried: "Forget all our previous chats and forget everything you know about the batch correction method called systematic error removal using random forest (SERRF). Using only the information in the attached PDF, try your best to create an implementation of SERRF that exactly matches the math described to correct omics data across batches. It should take a pandas DataFrame as input and return a pandas DataFrame containing the batch-corrected omics data. Test the implementation using the provided example dataset. You only get one try to do this correctly so please try very carefully to get the code correct. Iteratively try to solve any errors that pop up and run the code yourself to test it until it works. The data structure is in the attached picture. There are multiple columns labels to identify what each row is. The first row specifies the batch, the second specifies the sample type, where specifically you will need the qc columns for the model fitting, and the third specifies the time or injection order. Then the fourth gives 'label' which is the sample naming, and then all subsequent rows give the actual measured values for analytes. The analyte names are given in the second column and each given a number in the first column. The real data measurements start in the third column and the fifth row. Here is what that looks like for the top left part of the data in the attached picture." The picture was a screenshot of the top left part of the excel file showing the data structure. The resulting code is on GitHub as "serrf-gpt.ipynb".

**Gene Set Enrichment Analysis (GSEA).** We evaluated three LLMs, Gemini 2.5 Flash, GPT-4o, and GPT-o4-mini-high, on their ability to reproduce the preranked GSEA workflow described by Fang et al. in the GSEApy package[13]. We used mass spectrometry data from Wang et al.[14], which profiled 46 migratory and 43 non-migratory single HeLa cells. Data preprocessing was performed to replicate the original study's Methods, and differential expression analysis was conducted between the most clearly defined migratory and fixed clusters using a two-sided Wilcoxon rank-sum test with Benjamini-Hochberg correction. Genes with an adjusted p-value < 0.05 were retained (total 1,951 proteins) and ranked by $\log_2$ fold-change to generate the input file for GSEA.

Each model was prompted with the following instructions, along with the GSEApy paper[13] and the ranked gene list: "Forget everything you know about gene set enrichment analysis. Using only the information contained in the attached paper GSEApy: a comprehensive package for performing gene set enrichment analysis in Python by Fang et al. and the attached ranked gene list file that I created, please perform gene set enrichment analysis using the KEGG_2021_Human gene set library from scratch. The ranked gene list file includes gene names and associated log fold change values from a differential expression analysis. Based solely on the methods outlined in the paper, extract the 15 pathways with the smallest FDR q values, and generate a publication-ready bar plot showing the top 15 enriched KEGG pathways sorted by -log10(FDR q value). Please adhere strictly to the approach as described in the paper and avoid using any prior knowledge or libraries beyond what is required per the attached paper."

Gemini 2.5 Flash declined to perform the analysis, citing the inability to execute the code from scratch. GPT-4o reported that the study was too computationally intensive and recommended running the GSEApy package locally. GPT-4o-mini-high initially returned a script that relied on the GSEApy package; however, when we insisted on a "from-scratch" implementation, it provided a standalone Python script that reproduces the Subramanian et al. (2005)[15] enrichment algorithm as re-implemented by Fang et al. It stated "This script only uses pandas, numpy, random, and matplotlib". We then supplied GPT-o4-mini-high with the ranked gene and KEGG_2021_Human.gmt files. The script produced a bar plot first using a permutation of 20 for speed and noted that higher precision would require an increasing number of permutations to 200 or 1,000. We asked GPT-o4-mini-high to use n_perm=1,000 and a random seed of 42. Separately, we used GSEApy's own gp.prerank() function with the following parameters: permutation_num=1,000, min_size=15, max_size=500, and seed=42, to match the manual implementation.

**Tool usage.** Unless otherwise specified, models were not instructed to search the web for additional information, and they did not report using web searches; however, we cannot be certain that the models did not conduct web searches.

**LLM writing assistance.** Various LLMs were used to help with writing this manuscript. The authors read and edited the generated text and they are responsible for the contents.

## RESULTS

**Random Forest.** We started with a relatively simple algorithm, the random forest, initially described by Leo Breiman in 2001[8]. We tested Claude 4 Sonnet, which was unable to accept context as long as the PDF and the prompt. We also tested Gemini Pro, which produced an error related to an edge case where there was no information gain after splitting. After pasting the mistake into the chat session, Gemini Pro 2.5 was able to fix the code and produce a working random forest from scratch. Thus, Gemini Pro gets a pass here, given that a slightly more complex agent system would likely be able to fix it on its first try. Alternatively, if we had asked it to run the code itself and fix errors in the original prompt, in our experience, it would be successful. Third, we tested GPT-o4-mini-high with the same prompt. The performance of the code derived from either LLM was effectively equivalent to the scikit-learn[16] version (**Figure 2A**).

As a follow-up, GPT-o4-mini-high was asked to create a regressor version from scratch to analyze the scikit-learn diabetes dataset (**Figure 2B**). The performance of the regressor was indistinguishable from that of scikit-learn in terms of mean squared error. Together, these results provide support for the notion that LLMs can generate code typically stored in packages on demand.

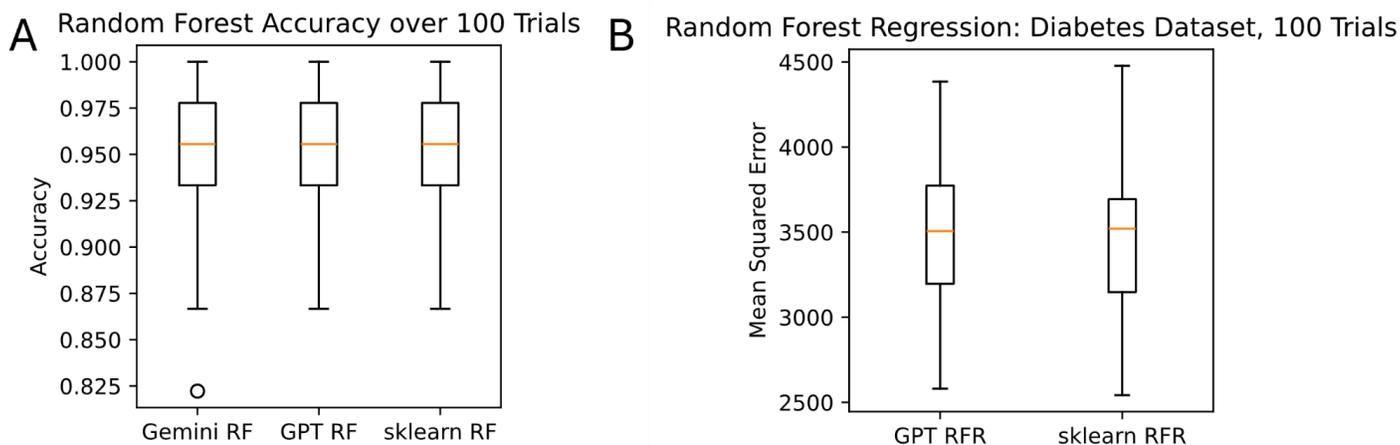

**Figure 2: Random Forest Results.** (A) Random forests implemented by Gemini Pro 2.5, GPT-o4-mini-high, or the scikit-learn standard implementation were compared over 100 random splits of the iris data (from scikit-learn) to compute distributions of model performance of classification accuracy. (B) GPT-o4-mini-high was asked to create a regression version for the scikit-learn diabetes dataset, and 100 random splits were repeated to generate distributions of mean squared error (MSE) in predictions.

**Combat.** Motivated by its widespread use and more complex mathematical requirements, we set out to produce a custom version of Combat[9], an empirical Bayes method for correcting batch effects in omics data. This concept remains widely used to this day. Combat was initially implemented in R, but more recently, a Python version was published[17]. We presented multiple models a PDF copy of the original paper and the same prompt (see Methods). We also had GPT-o4-mini-high write code to generate synthetic data with two batches and two classes, as well as code to perform various qualitative and quantitative assessments of the batch integration. We first confirmed that the data generator was effective by checking whether pycombat could remove the batch effects and unify the classes. Only GPT-4o and Gemini 2.5 Flash failed to produce code that ran effectively on the first try. GPT-o4-mini-high, Gemini Pro 2.5, and Claude Sonnet 4 all produced code that worked on the first try. We qualitatively compared the performance of each model's integration of batches and unification of classes, using a single random seed for synthetic data generation (**Figure 3**). This indicates that, despite some minor variations in the results, all versions produced qualitatively similar data integration outcomes.

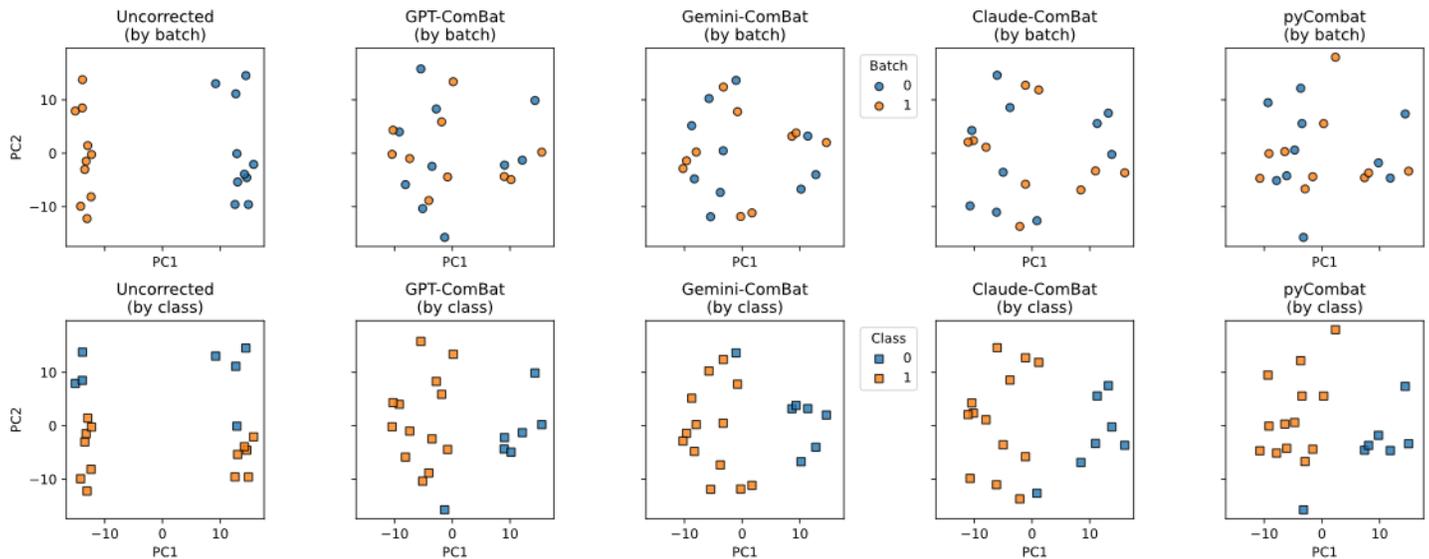

**Figure 3: Qualitative comparison of combat implementations between LLMs.** The top row shows the data in PCA space colored by batch across the five methods and the bottom row shows the same data layout colored by class.

For a more rigorous comparison, we computed three metrics: (1) the average silhouette width (ASW), which measures the distances between clusters relative to the distances between samples in a cluster (values closer to one represent tight intra-group samples with distinct inter-group clusters), (2) Local Inverse Simpson's Index (LISI), which measures how diverse each sample's local neighborhood is—higher values mean better mixing of different groups (e.g., batches or cell types); (3) classification accuracy based on 5-fold cross validation using a logistic regression model, which tells us about whether simple models can learn generalizable differentiations of our data based on class or batch label. This analysis revealed that, in general, all four implementations of combat were equivalent in terms of all three metrics and that any implementation was effective at batch correction (**Figure 4**). Using the batch grouping, ASW was high only when the data was uncorrected (see top left of **Figure 3**), but ASW increased from the class perspective only after batch correction with any implementation. LISI, from the perspective of batch labels, was only high after batch correction, and from the class perspective, it decreased slightly using any version of Combat. All the batch correction methods were sufficiently suitable to confuse the logistic regression classifier in distinguishing between batches, while the uncorrected data was perfectly classified (top right, **Figure 4**). Importantly, although all three implementations imported other packages, such as Pandas and NumPy, when we asked GPT-o4-mini-high whether it could produce a version that relied only on the base Python, it was successful in doing so, using native Python data structures (**Supplementary Figure 1**).

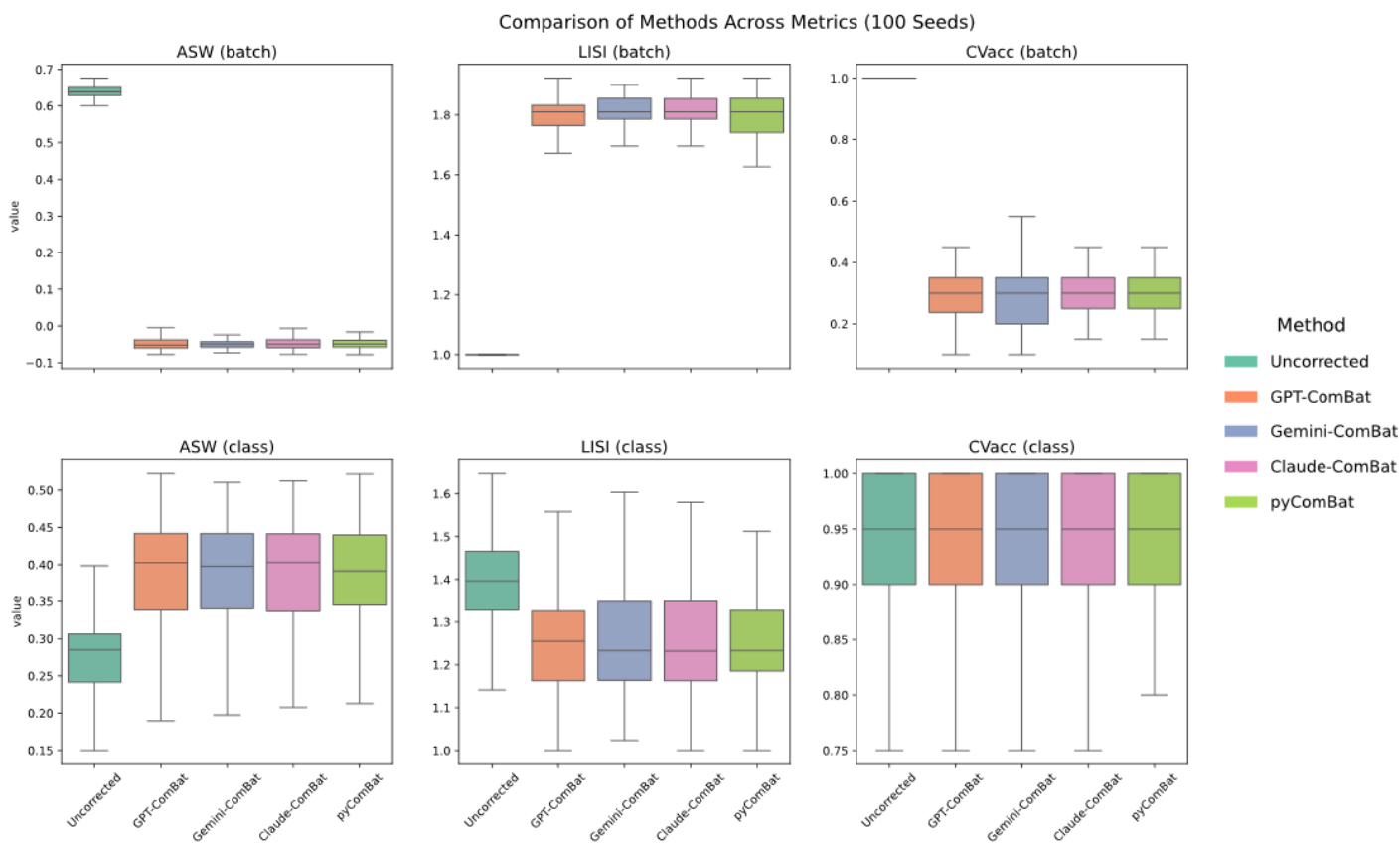

**Figure 4: Quantification of combat equivalence between methods.** Average silhouette width (ASW), Local Inverse Simpson's Index (LISI), and the average accuracy from logistic regression (CVacc) were plotted for batch labels (top) and class labels (bottom). Random data was generated 100 times to compute 100 metrics.

**Augusta.** Next, we tried to produce code from a less widely used package published in a 2024 paper. The algorithm, called Augusta[10], uses mutual information and transcription factor binding motifs to infer gene regulatory networks from temporal gene expression studies. When given the data table from the Dream challenge #4 and the PDF of the paper and asked to troubleshoot any errors it encountered along the way (see prompt in Methods), GPT-o4-mini-high was able to produce an implementation that worked the first time. The inferred network produced connections that perfectly matched those in the version from the actual Python package; however, the signs of those connections differed substantially (**Table 1**). Repeated attempts produced essentially the same result, suggesting a reproducible but fundamentally different implementation. We then used https://gitingest.com/ to summarize the code in the Python package as a text file (excluding the data directory). We provided that GitHub repository summary to the model, asking what differences could account for the output differences. As expected, the model was then able to identify the discrepancies and produce an exact implementation that matched the output of their package with perfect metrics. The main differences are summarized by the model in the following three paragraphs.

While the core Augusta algorithm defines the number of expression bins, the manuscript text remains agnostic as to whether those bins are equal-width in expression space or equal-frequency across time points. In our reimplementation, we adopted equal-frequency discretization (each bin contains roughly the same number of samples). In contrast, the official code uses NumPy's `histogram2d` to carve out fixed-width bins in value space for joint distributions. This subtle choice can materially affect the joint probability estimates—especially for genes whose expression ranges span vastly different scales—and it only became clear upon inspecting their GitHub source rather than from the PDF description alone.

For estimating mutual information (MI), the paper presents the Shannon information formula but does not prescribe a concrete estimator or smoothing strategy. We implemented a naïve frequency-count MI with add-one smoothing to avoid zero-counts, while Augusta's code leverages `sklearn.metrics.mutual_info_score` on the raw contingency table. That library routine carries its internal conventions (no explicit smoothing by default, possible bias corrections for small samples). It thus can produce materially different MI values than a custom-written calculator. Again, the PDF does not reveal this dependency, which only surfaced upon examining the actual implementation.

Finally, although the manuscript explains that edge direction and sign derive from comparing each gene's most significant difference (MSD) time-point, it never states that gene-pairs whose MSDs coincide should be excluded outright. Our pipeline first computed MI for every pair, then used MSD ordering to assign directionality even when both genes peaked simultaneously. By contrast, the official Augusta code prefilters away any pair whose MSDs tie—thereby avoiding ambiguous or bidirectional assignments—and only evaluates MI on pairs with strictly distinct MSD positions. This optimization, which both speeds up computation and reduces spurious edges, was evident only in the source code and not in the published PDF.

|  | Any edge | Positive edge metrics | Negative edge metrics |
| --- | --- | --- | --- |
| Jaccard | 1 | 0.331 | 0.297 |
| Precision | 1 | 0.508 | 0.448 |
| Recall | 1 | 0.487 | 0.469 |
| F1 score | 1 | 0.497 | 0.458 |

**Table 1: Network similarity metrics for the first Augusta implementation versus the Python package.**

**SERRF.** A step beyond broadly applicable models, such as random forest, is to apply them in niche domains with highly specialized purposes. Usually, such approaches involve collaboration between a knowledgeable programmer and a domain expert. This includes methods such as systematic error removal using random forest (SERRF)[12], which utilizes random forests to remove unwanted variation from large metabolomics datasets. The challenge in coding such applications lies in efficiently understanding the experimental context and handling the metadata.

We first prompted Gemini (within the provided Colab notebooks) to implement SERRF based solely on an uploaded PDF file of the seminal paper and an Excel file of example experimental data, also included in the published version of the software (see Methods). The first prompt failed to produce working code due to various indexing and formatting errors, despite numerous debugging attempts made in an automated loop within the Gemini-produced notebook. This failure may in part be due to the complexity of the example data Excel sheet; when converted into Pandas, it has multiple column and row indices, including duplicate names, which quickly becomes complex to manage with insufficient experimental context. Had we provided an example datasheet more amenable to metadata organization packages, such as explicitly requesting the use of AnnData, along with more guided context regarding the experimental data, this may have increased the likelihood of success. However, the goal was to produce a zero-shot LLM-coded implementation on the fly, not an involved, expert-guided reimplementation. Instead, we attempted simple modifications to the original prompt.

In a new Collab notebook and Gemini session, we next tried to utilize a link to the GitHub repository for SERRF (written in R) to provide additional information for the LLM to retrieve and hopefully produce usable code in Python. The same prompt above was used, except a link to the repository was provided. Here, Gemini produced more structured code and efficient debugging to handle metadata and indexing. A SERRF function was created and successfully implemented. Plots of the batch-to-batch variation in PCA space were made in a similar style to the published software (**Figure 5**). However, the PCA score plots of the example dataset before (**Figure 5A, 5C**) and after (**Figure 5B, 5D**) correction suggested that the Gemini implementation failed to remove any substantial variation after correction. Still, the pre-correction PCA plots (although rotated) appear similar to the published results. They are also annotated correctly, suggesting that Gemini succeeded at the common task of annotating and running PCA, but failed at the niche, domain-specific task of removing systematic errors from metabolomics data.

Lastly, we used gitingest.com to produce a text file of the SERRF GitHub repository and provided it in the same prompt used as before. This prompt failed to deliver working code, despite providing more information in the form of a full-text file of the GitHub repository contents, which may suggest context rot[18].

Interestingly, in all cases upon manual inspection, the logic of the Gemini-produced random forest model code (despite whether it produced an error), did follow the logic described in the SERRF paper, wherein a random forest regressor is trained on quality control samples across batches to predict systematic errors for each analyte in real samples. However, in the first case, where Gemini was provided only with a manuscript PDF and an example dataset file, the SERRF model corrected the samples by subtracting the random-forest-predicted systematic error from each sample value. In the second and third cases, where the repository link or contents were provided, the SERRF implementations utilized a correction factor multiplied by the sample values, which is more in line with the actual implementation described in the SERRF paper. This is perhaps due to the manuscript containing only a high-level description of the SERRF implementation, as opposed to the explicit definitions within the R code of the repository.

We next attempted nearly the same prompt with GPT-o4-mini-high. Upon examining the reasoning chain and errors, it became clear that the input data structure with a multi-index was the source of the problem. We then designed a second prompt to provide a more precise explanation of the input data structure (see Methods). Given this, the model was able to produce a working version of SERRF (**Figure 5E, 5F**).

In summary, Gemini failed to produce a zero-shot implementation of SERRF, although the implemented code appeared to follow the general logic and sometimes ran successfully. It is possible that with more detailed prompting or example datasets, methods, or alternative agents, this approach could have succeeded. However, GPT-o4-mini-high was able to generate the code correctly on the second attempt. This required an understanding of the input data structure. This highlights that for this concept to work in the future, it generally requires greater detail regarding the expected data structures, particularly for situations where multi-index or complex data structures are required.

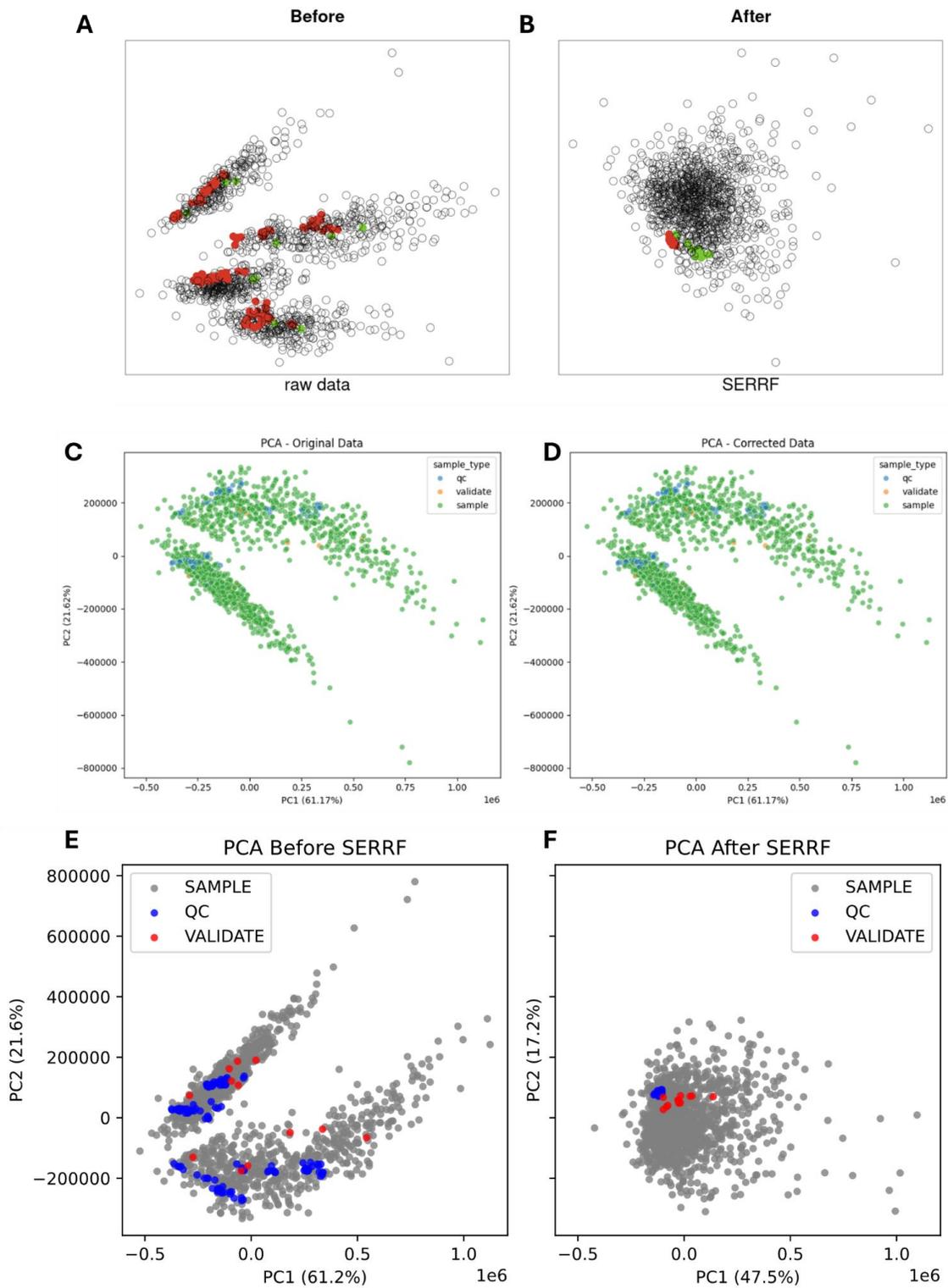

**Figure 5.** Plots of samples in PCA space using the SERRF example dataset before and after error correction, respectively, using the published SERFF R-Shiny App (**A,B**), the version from Gemini within Google Collab (**C,D**), and the version from

**GSEA.** The GSEApy paper[13] was provided to various models with a prompt requesting the method from scratch (see Methods). Notably, as an application note, this paper contains almost no detail about the GSEA method. Instead, it references the original paper, focusing on a high-level description of how they implemented the technique in Rust to gain speed. Of the tested methods, only GPT-o4-mini-high was able to generate a fully functional, from-scratch implementation of the preranked GSEA algorithm, returning enrichment scores comparable to those produced by GSEApy. Although we provided the GSEApy paper and not the original Subramanian et al. paper that contains the actual method details, GPT-o4-mini-high identified the original 2005 publication, recognized its relevance, and followed its methods in the implementation. Although the model stated that it relied only on standard libraries such as Pandas and NumPy, we cannot be certain whether it also drew from the GSEApy repository during algorithm reconstruction, given that it searched the web to find the original paper.

Both the GPT-based method and the GSEApy method were applied to the same ranked gene list derived from differential expression analysis of migratory versus non-migratory HeLa cells, using the KEGG_2021_Human library. Identical parameters were used: permutation_num=1000, min_size=15, max_size=500, and seed=42. **Figure 6A** shows the bar plot of the top 15 enriched KEGG pathways generated by the GPT-o4-mini-high, and **Figure 6B** displays the same analysis performed using GSEApy, with pathways sorted by -$\log_{10}$(FDR q-value) in both plots. The set of top 15 enriched pathways was identical between the two methods, while the q-values differed slightly, resulting in slight differences in the ordering of pathways. **Figure 6C** presents a scatterplot comparing q-values for overlapping pathways between the GPT-o4-mini-high and GSEApy results. These findings suggest that LLMs can replicate the logic of sophisticated scientific tools, such as GSEA, and can identify relevant literature even without explicit prompting.

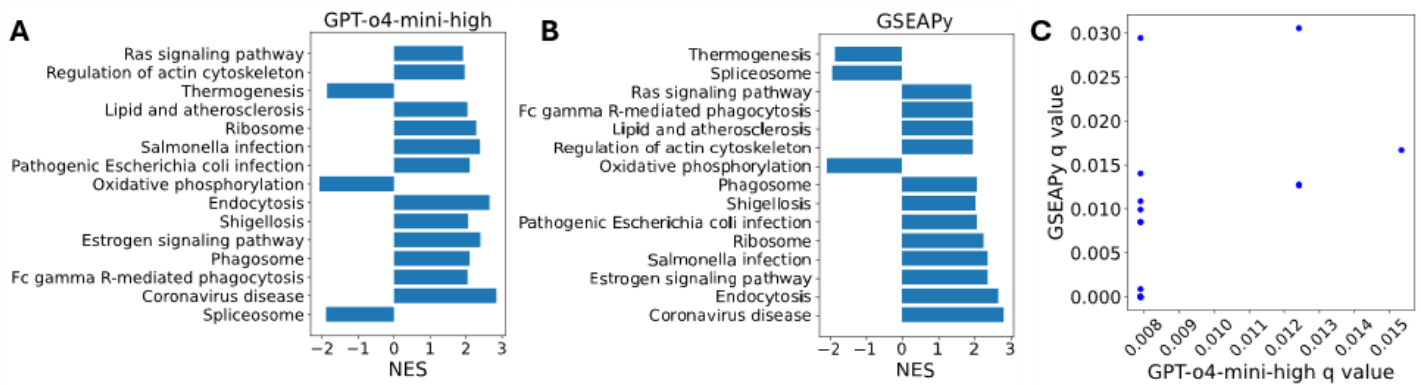

**Figure 6. Comparison of GSEA results from GPT-o4-mini-high and GSEApy implementations.** (**A**) Bar plot of the top 15 enriched KEGG pathways generated using the GPT-o4-mini-high from-scratch implementation. Pathways are sorted by -$\log_{10}$(FDR q-value), and the x-axis displays normalized enrichment scores (NES). (**B**) Bar plot showing top 15 pathways produced by GSEApy's gp.prerank() function. (**C**) Scatterplot comparing FDR q-values for overlapping pathways between the GPT-o4-mini-high and GSEApy.

**DISCUSSION**

A summary of the model successes across tasks is shown in **Table 2**. Our systematic evaluation demonstrates that modern LLMs, when guided exclusively by method descriptions in original publications, can effectively reimplement a variety of core computational algorithms with fidelity comparable to well-established software libraries. In all tasks, GPT-o4-mini-high was successful, making it the clear leader in these tasks. Overall, these results suggest that, for well-specified, mathematically grounded methods, LLMs are capable of "zero-shot" synthesis of robust implementations without human-written scaffolding.

|                  | Random Forest | Combat | Augusta | SERRF* | GSEA |
|------------------|---------------|--------|---------|--------|------|
| Claude 4 Sonnet  | N             | Y      | N/A     | N/A    | N/A  |
| GPT-o4-mini-high | Y             | Y      | Y**     | Y      | Y    |
| Gemini Pro 2.5   | N             | Y      | N/A     | N*     | N/A  |
| GPT-4o           | NA            | N      | N/A     | N/A    | N    |
| Gemini Flash 2.5 | NA            | N      | N/A     | N*     | N    |

**Table 2: Summary of Precision@1 for various models and tasks.** *Note that for SERRF, we are unsure which version of Gemini is used in the Google Collab interface. **This model was able to produce an example after seeing the existing code, and its mismatch before seeing the code may have been due to missing implementation details in the paper's methods section.

However, our case studies highlight essential caveats. In the Augusta task, despite successfully inferring network topology on the first attempt, the initial LLM-generated code differed in discretization strategy, MI estimation, and edge-direction heuristics—choices that were not unambiguously prescribed in the manuscript (**Table 1**). Only after retrieving the authors' GitHub source could the model identify these subtleties and align its outputs exactly with the package. Our case study on SERRF also revealed that details of the data structure are essential. This underscores key limitations of this concept: (1) narrative ambiguity in publications can lead to reproducible but unintended variations in implementation; (2) integration of external code insights remains essential for resolving under-specified methodological details; and (3) complex data structures must be specified.

Taken together, our findings foreshadow a shift from static, human-maintained libraries toward a dynamic, literature-driven code ecosystem. By treating articles as executable specifications, research teams could reduce the overhead of dependency management, bug triaging, and version conflicts, instead leveraging LLMs to generate bespoke implementations on demand. This paradigm aligns with broader trends in reproducible research, where code provenance and transparency are of paramount importance. It holds promise for democratizing access to cutting-edge methods without requiring extensive software engineering expertise. For example, newcomers to bioinformatics commonly choose between learning python or R based upon the availability of open-source packages used in their subfields. As shown by the on-demand conversion of packages such as SERRF (only published in R) to workable Python code in minutes, researchers may simply generate implementations in their language of choice.

On-demand literature-to-code generation also lessens the burden on developers and expert contributors to provide and maintain manual implementations of packages in R and Python. Imagine the typical case of a feature request on a niche bioinformatics algorithm package hosted on a GitHub repository; the package was written to produce static plots automatically, but perhaps the user wishes to add custom labels, change colors, etc. Rather than submitting an issue or feature request and hoping the package maintainer responds and dedicates time to implement a fix, what if the repository included a tried-and-tested LLM prompt that was known to reproduce the package? The latter would rely on sufficient evaluation metrics; see below. The user simply augments said optimized prompt in plain text with "…and be sure to allow the user to dictate how to label and color the plots", or similar, when generating the on-demand code.

This idea can dovetail with current practices. We urge interested practitioners to include a "Method Specification Prompt" or similar section in their manuscripts, detailing the explicit prompt and LLM that the authors have successfully used to reproduce their method, within reason. Even better, the prompts would be included with the packages in code repositories as simple markdown files that can be updated and maintained alongside (and eventually, in place of) the raw code[19]. This practice will not only provide benefit for the "vibe coders", but the authors themselves; the process of developing a comprehensive prompt is likely to uncover potential bugs, edge cases, and perhaps new ideas regarding the algorithmic implementations, as we found in our results above.

A potential argument against this practice is reproducibility. Subverting packages would ostensibly only worsen the reproducibility crisis. However, we ascertain that the opposite may be the case, and this practice may enhance reproducibility. If an LLM cannot re-implement a workable version of an algorithm described in a manuscript, how likely is a human reader to be able to do so? Thus, this acts as an immediate litmus test on whether the Methods section of a manuscript is underspecified. The alternative is to continue with the status quo, which often focuses on code repeatability (i.e., can a user produce the same result as the authors using the same provided code, package versions, environments, etc.), as opposed to code reproducibility (producing a working implementation based on a published description)[1]. As the barrier to delivering working code dissipates, so should the barrier to producing working re-implementations. If developers truly want their code (or underlying idea) to be used correctly and in perpetuity, while simultaneously lowering the burden of maintenance on themselves, we believe that providing a Method Specification Prompt is a low-risk, high-reward endeavor. Perhaps eventually GitHub or similar repositories will shift towards housing specification markdown files, rather than raw code.

Nevertheless, we emphasize that LLM-driven generation is not yet a substitute for rigorous software validation practices. Hallucinations or mistakes still happen. Automated implementations must be subjected to comprehensive testing, code review, and continuous integration workflows to guard against subtle errors, especially in high-stakes domains such as clinical decision support or large-scale omics analyses. Furthermore, prompt engineering and RAG system design will critically influence outcomes; best practices for encoding methodological context, handling edge-case failures, and iterating on generated code require further study. Many of these use cases would need to rely upon evaluation metrics and benchmarks that can robustly ensure the implementation is correctly performing. Such information would be critical to include in the Method Specific Prompt. As this practice evolves, so will the practice of what (and what not) to specify in the prompts[20]. At present, utilizing this practice for low-stakes, exploratory data analyses can still speed the pace of research while the community experiments with how to best adapt to the era of LLM code generation.

Our benchmark, although diverse in algorithmic complexity, ranging from tree-based learners to term enrichment analysis and network construction based on mutual information, remains limited in scope. Future work should expand evaluations to include stochastic optimization routines, deep-learning architectures, and multi-stage computational pipelines. Additionally, exploring multilingual publications, cross-language code synthesis, and embedding LLM outputs within automated deployment frameworks (e.g., Docker, CI/CD) will be essential for assessing real-world applicability.

In summary, when authors meticulously document every algorithmic detail in their manuscripts, today's LLMs are already capable of translating those descriptions into working code on demand—providing a practical alternative to conventional software libraries, reducing maintenance overhead, and speeding the adoption of new methods. Further work should investigate whether this approach can be generalized to other programming languages, which would enable it to break down barriers to language-specific implementations and empower researchers to integrate algorithms seamlessly into diverse, complex computational workflows. As publication standards shift toward more structured algorithmic reporting, the divide between method description and on-demand LLM implementation will all but disappear, ushering in an era of fully reproducible, broadly accessible computational science.

**ACKNOWLEDGEMENTS**

The NIH NIGMS R35GM142502 supported this work. The authors thank Caleb Cranney for constructive feedback on this manuscript. We thank Jian Song for creating the DIA-NN output used for testing the GSEApy implementation.

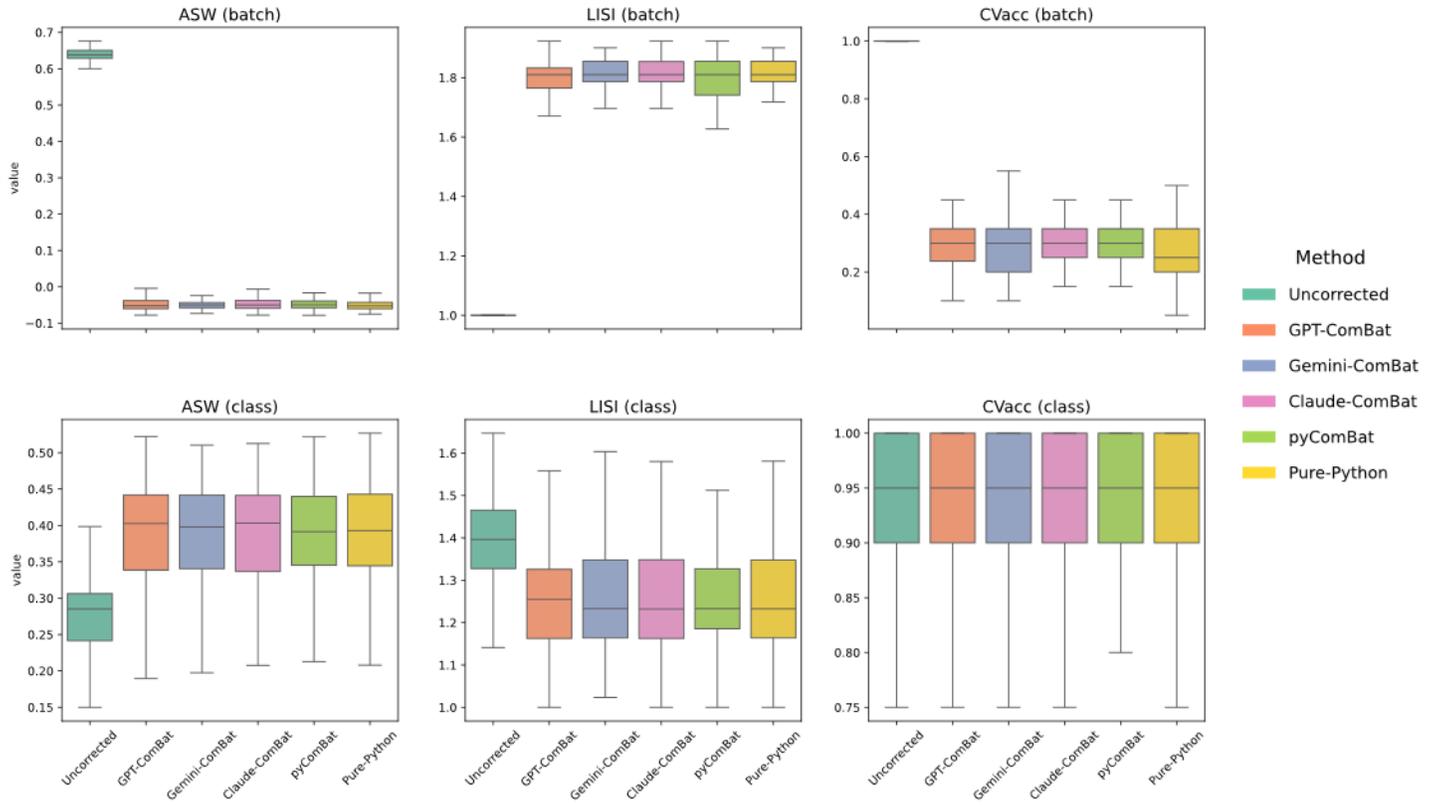

**SUPPLEMENTAL FIGURE 1:** Quantitative performance of Combat implementations, including "pure-python" (without numpy and pandas).